\documentclass[a4paper,12pt,leqno]{article}

\usepackage[margin=2.54cm]{geometry}

\usepackage{tikz}
\usepackage{amsfonts, amsmath, amssymb}  
\usepackage{times}
\usepackage{tablefootnote}
\usepackage[bottom, hang, flushmargin]{footmisc}
\usepackage{multirow}
\usepackage{xcolor}

\usepackage{natbib}

\usepackage{verbatim}
\usepackage{caption}
\usepackage{subcaption}
\usepackage{booktabs, array}
\usepackage{geometry}
\usepackage{colortbl, arydshln}
\usepackage[english]{babel}
\PassOptionsToPackage{hyphens}{url}

\usepackage[ruled]{algorithm2e}

\usepackage{hyperref}
\usepackage{cleveref}
\usepackage{chngcntr}

\usepackage{float}
\usepackage{titlesec}
\usepackage{graphicx}
\definecolor{mygray}{gray}{0.9}
\usepackage{mathabx}
\usepackage{mathtools}
\DeclarePairedDelimiter\ceil{\lceil}{\rceil}

\addto\extrasenglish{%
}

\DeclareGraphicsExtensions{.pdf,.png,.jpg,.bmp, .tif}

\usepackage[toc,page]{appendix}
\usepackage{etoolbox}

\AtBeginEnvironment{tabular}{\singlespacing}
\AtBeginEnvironment{verbatim}{\singlespacing}

\newcommand{\e}{\text{e}}
\newcommand{\ssb}{\textit{SSB}}
\newcommand{\ssbnaught}{\textit{SSB0}}
\newcommand{\sr}{\text{SR}}

\newcommand{\glinf}{\left( 57.08, 77.70 \right)^\intercal }
\newcommand{\gbeta}{\left( 1.09, 0.79 \right)^\intercal }
\newcommand{\gsigma}{\left( 0.15, 0.15 \right)^\intercal }
\newcommand{\alphaM}{\left( 0.79, 0.42 \right)^\intercal}

\newcommand{\rhoIota}{0.75}
\newcommand{\muIota}{-2}
\newcommand{\sigmaIota}{1}
\newcommand{\kappaEta}{\exp\left(3\right)}
\newcommand{\tauEta}{0.1}
\newcommand{\rhoEta}{0.75}
\newcommand{\psiEta}{0.75}
\newcommand{\Flim}{(1\mathrm{e}6, 1\mathrm{e}6)^\intercal}
\newcommand{\Fint}{1.5}
\newcommand{\fone}{(0,0)^\intercal}
\newcommand{\ptarg}{0.90}	
\newcommand{\Finttwo}{1.0}
\newcommand{\Fintthree}{0.5}
\newcommand{\kappaEps}{\exp\left(5\right)}
\newcommand{\tauEps}{0.0005}
\newcommand{\rhoEps}{0.75}
\newcommand{\rcut}{(0.05, 0.05)^\intercal}
\newcommand{\muEps}{-7}

\newcommand{\rNaught}{(\exp(7.0), \exp(4.5))^\intercal}
\newcommand{\rrange}{(0.5, 0.5)^\intercal}
\newcommand{\steepness}{(0.9, 0.9)^\intercal}

\usepackage[doublespacing]{setspace}

\begin{document}

\begin{flushleft}
	{\LARGE\bf
		spatialSim: multi-species spatiotemporal size-structured operating model for management strategy evaluation }\\[10mm]
	\baselineskip=6mm
	{\large\bf
		Christopher D. Nottingham and Russell B. Millar} \\
	Department of Statistics, 
	University of Auckland,
	Private Bag 92019,
	Auckland, New Zealand \\[3mm]
\end{flushleft}
\hrulefill

\noindent {\bf Corresponding author}: Christopher Nottingham
(e-mail: c.nottingham@auckland.ac.nz).

\newpage
\begin{abstract}

Spatiotemporal processes have the potential to be one of the most influential factors governing how fisheries targeting sedentary species respond to harvesting. Despite this, management strategy evaluation often fails to account for space or does so at low resolutions due to compute constraints. In this paper, a multi-species spatiotemporal size-structured operating model for sedentary species is presented. The model combines a spatially continuous Gaussian Markov Random Field model of the population dynamics with an areal harvesting model that supports preferential targeting and site selection constraints (e.g., economic constraints). This approach is very compute efficient, which makes it feasible to simulate realistic fisher dynamics and catch data at true spatial scale (e.g., the swept area of a dredge). The New Zealand surfclam fishery was used as a case study to demonstrate the versatility of the operating model and to showcase the simulation of localized depletion, which was manifest in the generation of realistic catch-per-unit-effort data that were uncorrelated with the trends in population abundance. The model is available as part of the open-source R package spatialSim.

\end{abstract}


\section*{Introduction}
Determining an optimal management strategy is a critical part of sustainably managing a fishery. Traditionally, management advice has been based on a `best assessment' of the resource, which involves fitting a model to fisheries data and evaluating uncertainty using confidence intervals and sensitivity tests. Recommendations for management action are then determined by applying a harvest control rule to values derived from the model or from model-based projections that assume a constant catch or fishing mortality rate \citep{but07}.

Contemporary fisheries management has started moving away from the traditional approach in favor of a more holistic simulation procedure known as management strategy evaluation (MSE) \citep{had12}. MSE has its origins in the field of operations research \citep{cha95}, which is centered around problem-solving methodology that can be used in the pursuit of improved decision-making and efficiency. The MSE process uses a feedback loop to compare the performance of alternate fisheries management strategies with respect to achieving pre-assigned management objectives. The loop contains a simulation model known as the operating model (OM) that represents the true state of nature and a management strategy, which is made up of an estimation method (e.g., a stock assessment model) and a harvest control rule. The sequence of events in the feedback loop are: (1) simulate the true state of nature from the OM for assessment period $t$, (2) simulate fisheries data (e.g., commercial catch and survey data) from the OM, (3) estimate the stock status from the assessment model using the simulated data, (4) apply the harvest control rule to determine management recommendations (e.g., quota, effort limit, size limit or time-area closure) for assessment period $t+1$, and (5) subject the management recommendations to implementation error such as quota overages caused by unreported catch. The loop is repeated for a sufficient number of future assessment periods and performance statistics are calculated. 


MSE can be a very valuable tool for fisheries management. However, it relies heavily on the specification of the OM(s) and, in particular, the range of assumptions and uncertainties that are included in the modelled processes. In the literature, there are a large number of cases where the adequacy of the assumptions are questionable. 
One widespread issue is a ``home cooking" problem, where the OM used to test a management strategy or assessment model is developed by the same person and makes the same or similar assumptions as the assessment model. This approach is suboptimal as it is likely to result in an unfair test, particularly when evaluating assessment methods. 
Another issue that is also prevalent in many cases and pertinent in the context of this work is the inadequate representation of spatial processes, which are often ignored \citep{pun16a} or incorporated by modelling a stock across a relatively small number of discrete zones \citep[e.g.,][]{pun09, fay11}. 
Generally, this means the effects of localized depletion are also ignored or poorly captured in the simulated system and catch data.
Consequently, the performance of any assessed management strategy will likely be overstated. This issue has the potential to lead to very misleading management recommendations, particularly for sedentary species that exhibit a significant amount of spatially heterogeneity, density dependence, and limited population connectivity.

In this paper a novel multi-species OM with spatiotemporal size-structured dynamics is presented. It provides scientists with a tool that can be used to avoid the ``home cooking" problem and has a bespoke formulation balancing simplicity and complexity for simulating high resolution spatiotemporal dynamics of correlated populations that exhibit spatially heterogeneity, density dependence, and limited population connectivity. The model provides users with large amounts of flexibility when specifying systems, and it was developed in a modular way to facilitate the further development of structures and dynamics that are currently not supported.


The model's state dynamics are an extension of the single-species non-spatial size-structured dynamics described by \citet{pun13}. They incorporate spatiotemporal variation in recruitment and natural mortality using Gaussian Markov Random Fields (GMRF), which results in the simulated populations varying continuously across space. The spatially continuous population dynamics are combined with an areal model that can be used to simulate scientific surveys and commercial fishing on a grid containing the projected numbers-at-size.

The harvest model supports the specification of different levels of preferential targeting (bias to harvest from areas with higher densities of individuals) and different levels of fishing intensity across space. In addition, it supports user specified site-selection probabilities that can be used to incorporate factors such as economic constraints and area closures. Combining the continuous population dynamics with the areal harvesting model is computationally very efficient, which makes it feasible to simulate catch data to scale (e.g., the swept area of a dredge or diver).

The model also supports density dependence in recruitment through the incorporation of a spatially explicit stock-recruit function, where the number of recruits at a given location depend on the density of sexually mature individuals within a user-set spatial range of the location. This specification makes the assumption that the simulated population(s) self-recruit with no input from outside the spatial domain. Therefore, it is only appropriate for species/populations that do not exhibit significant larval drift, for which, there are likely to be many \citep{han14, tes16}.

The multi-species aspect of the model provides users with the option of specifying a species correlation structure on recruitment and assumes that the natural mortality rates of each species are proportionally related across spacetime. This allows fishery systems comprising species with correlated spatiotemporal population structures to be simulated, and facilitates a straightforward way of comparing multi-species assessment methods for fisheries with these characteristics. This part of the model was developed to be a simple as possible, while retaining the complexity required to answer a set of targeted questions and produce insightful results. This distinguishes this model from a lot of the multi-species alternatives \citep[e.g.,][]{chi05, gra06, ful11}, which are often overly complex, slow to run, and difficult to specify/understand.

The model is coded in the C++ template language TMB \citep{kri16}, which has a number of useful features. These include an elegant interface with R, simulation routines for GMRFs, multidimensional arrays, ragged arrays, sparse matrices and fast linear algebra routines through the Eigen library. In the presented work, automatic differentiation is not required and, therefore, double data types are used in place of CppAD data types. This enables a wider range of C++ functionality, which is taken advantage of through C++11 routines. This work is an open-source project and is available in the R \citep{rcor20} package, spatialSim (\url{https://github.com/cnotti/spatialSim}).






The following sections of this paper first outline the structure of the operating model and different options users have when specifying each of the simulated processes. These include the growth of individuals, natural mortality, harvesting, recruitment, and model initialization. Following this, a simple case study of the New Zealand surfclam fishery is presented. The case study demonstrates the versatility of the operating model and the importance of incorporating spatial processes into a MSE. In addition, it provides insight into the computational cost of running the model by presenting execution times for a number of different sized spatial domains.

\section*{Model overview}
The model is made up of two phases, which include an initialization phase, during which the populations are initialized, and a harvesting phase that represents the period in which harvesting occurs.  Both phases are centered around a state equation that describes how the numbers-at-size in the simulated populations evolve through spacetime. Processes that are part of the state dynamics include the growth of individuals, natural mortality, harvesting, and recruitment. The dimensions of the different structures of the model are controlled by a number of user-set parameters. These are presented in \autoref{tab_n}.

\subsection*{State dynamics}
The state equation is loosely based on the size-structured dynamics described by \citet{pun13}, with the addition of multi-species processes that describe spatial and temporal variation in recruitment and natural mortality. The spatial processes are modelled using the stochastic partial differential equation (SPDE) approach of \citet{lin11}. Hence, the numbers-at-size vary continuously across the spatial domain of the fishery $\Omega$, which is defined by an isometric triangulation with $v_S$ nodes, where each node represents a location referenced by $s$,
\begin{equation}
	\label{eq_N}
	n_{c,s,t+1,j} = \sum_{k} G_{c,j,k}  \left( n_{c,s,t,k} - x_{c,s,t,k} + r_{c,s,t,k} \right)\e^{ - M_{c,s,t,k} }.
\end{equation}
Here, $n_{c,s,t,k}$ is the number of individuals belonging to species $c$ that have size in the interval $Q_{c,k} = (l_{c,k-1}, l_{c,k})$ at time $t$ and location $s$, $G_{c,j,k}$ is the probability of an individual belonging to species $c$ growing from size interval $Q_{c,k}$ into size interval $Q_{c,j}$, $M_{c,s,t,k}$ is the instantaneous rate of natural mortality for individuals belonging to species $c$ with size $l \in Q_{c,k}$ at location $s$ between time $t$ and $t+1$, $x_{c,s,t,k}$ is the number of harvested individuals belonging to species $c$ with size $l \in Q_{c,k}$ at location $s$ between time $t$ and $t+1$, and $r_{c,s,t,k}$ is the number of individuals belonging to species $c$ with size $l \in Q_{c,k}$ that are recruited to the population at location $s$ between time $t$ and $t+1$.

\begin{table}[H]\centering
	\begin{tabular*}{\textwidth}{c p{11cm}}
		
		\multirow{1}{*}{ }
		\multirow{1}{*}{Parameter} & \multicolumn{1}{c}{\multirow{1}{*}{Description}}\\
		\midrule[1 pt]
		
		$v_B$ & Number of time periods in the initialization phase\\ 
		$v_Y$ & Number of years in the harvesting phase\\ 
		$v_P$ & Number of annual periods \\
		$v_C$ & Number of species \\
		$v_S$ & Number of triangulation nodes in spatial domain $\Omega$ \\
		$v_L$ & Number of size intervals \\
		$v_F$ & Number of cells in the harvest grid\\
		
		\midrule[1 pt]
	\end{tabular*}
	\caption{Parameters specifying the dimensions of model structure.}
	\label{tab_n}
\end{table}

\subsubsection*{Growth}
The probability of an individual transitioning from size interval $Q_{c,k}$ to $Q_{c,j}$ depends on the distribution of sizes within interval $Q_{c,k}$ and the distribution of growth conditional on those sizes, denoted $f_{U c}(u)$ and $f_{\Delta c}(\delta; u)$, respectively. It follows that 
\begin{equation}
	\label{eq_G}
	G_{c,j,k} = \int_{l_{c,j-1}}^{l_{c,j}} \int_{l_{c,k-1}}^{l_{c,k}} f_{\Delta c}(y - u; u) f_{U c}(u) \mathrm{d}u\mathrm{d}y.
\end{equation}
where $\delta = y - u$ is the growth of an individual between time $t$ and $t+1$, given initial size $u$ at time $t$. The two-dimensional integral in (\ref{eq_G}) is approximated using the method described by \citet{mil19}. This approximation assumes that the distribution of growth conditional on an individual's initial size is lognormally distributed and the sizes of individuals are uniformly distributed within each interval, but, for tractability, that individuals with initial sizes in the interval $Q_{c,k}$ have the same distribution for the expected amount of growth, $\delta_{c,k}$. The specification of $\mathbf{G}$ depends on a user-defined growth function and a parameter controlling the variability of $\delta_{c,k}$. Additionally, users must specify a function that defines how an individual's weight changes with growth. A detailed overview of the user-set growth parameters and functions is presented in \autoref{tab_growth}.

\begin{table}[H]\centering
	\begin{tabular*}{\textwidth}{c p{11cm}}
		
		\multirow{1}{*}{ }
		\multirow{1}{*}{Parameter/} & \multicolumn{1}{c}{\multirow{2}{*}{Description}}\\
		\multirow{1}{*}{Function} & \\
		\midrule[1 pt]
		$\mathcal{F}^w(l; \boldsymbol{\theta}_c)$ & Function determining the expected weight of an individual given its size $l$ and parameters $\boldsymbol{\theta}_c$, for each species $c \in \{ 1,...,v_C \}$\\
		$\mathcal{F}^\Delta(l; \boldsymbol{\theta}_c)$ & Function determining the expected growth of individuals given an initial size $l$, parameters $\boldsymbol{\theta}_c$ and $\Delta t$, for each species $c \in \{ 1,...,v_C \}$ \\	
		$\sigma^\delta_c$ & The sd of the growth increment on the log scale, for each species $c \in \{ 1,...,v_C \}$\\
		\midrule[1 pt]
	\end{tabular*}
	\caption{Parameters and functions that specify the growth of individuals.}
	\label{tab_growth}
\end{table}

\subsubsection*{Natural Mortality}
Natural mortality is specified using a parameter representing the expected instantaneous mortality rate for each species, $\boldsymbol{\alpha}_M$, and two stochastic processes that include a spatiotemporal field $\boldsymbol{\eta}$ and a temporally correlated process $\boldsymbol{\iota}$. In total, it depends on eight user-specified parameters, which are presented in \autoref{tab_M}. Explicitly, the instantaneous natural mortality rate on individuals belonging to species $c$ with size $l \in Q_{c,k}$ at location $s$ between time $t$ and $t+1$ is defined as,
\begin{equation}
	\label{eq:M}
	\Delta t M_{c,s,t,k} = \alpha_{M c}  \frac{\mathrm{logistic}(\iota_{t}) \exp(\eta_{s,t,k})}{\nu_M},
\end{equation}
where $\alpha_{M c}$ is the expected instantaneous mortality rate for species $c$, $\nu_M$ is a normalizing constant that standardizes the fraction to have a mean of one, the exponential term ensures that spatiotemporal variation in natural mortality is on the interval $(0, \infty)$, the logistic term provides additional flexibility by allowing temporally auto-correlated scaling of the spatiotemporal field (e.g., cyclical scaling), while retaining stability by bounding the term to be on the interval $(0,1)$, $\iota_{t} \sim \mathrm{N}(\mu_\iota(1- \rho_\iota) + \rho_\iota \iota_{t-1}, \sigma_{\iota})$ is a first order autoregressive (AR1) temporal process, and $\boldsymbol{\eta}$ is a Gaussian stochastic process with correlation between space, time, and the sizes of individuals, $\mathrm{vec}(\boldsymbol{\eta}) \sim \mathrm{MVN}(\mathbf{1}\mu_\eta, \sigma_\eta \mathbf{S}_\eta \otimes \mathbf{T}_\eta \otimes \mathbf{L}_\eta)$. Here, $\mathrm{vec}$ denotes the vectorization operator, which stacks the elements of an object over its dimensions to form a vector, $\otimes$ is the Kronecker product, $\mathbf{1}$ is a column vector of ones, $\mu_\eta$ is the marginal mean of $\boldsymbol{\eta}$, $\sigma_\eta$ is the marginal sd of $\boldsymbol{\eta}$, and $\mathbf{S}_\eta$, $\mathbf{T}_\eta$, and $\mathbf{L}_\eta$ are the correlation matrices for space, time, and size, respectively. The spatial, temporal, and size components of $\boldsymbol{\eta}$ are separable, such that, if time and size are marginalized, $\boldsymbol{\eta} \sim \mathrm{MVN}(\mathbf{1}\mu_\eta, \sigma_\eta \mathbf{S}_\eta)$ is a Mat\`ern field, if space and size are marginalized, $\eta_t \sim \mathrm{N}(\mu_\eta(1- \rho_\eta) + \rho_\eta \eta_{t-1}, \sigma_\eta)$ is an AR1 process, and if time and space are marginalized, $\eta_k \sim \mathrm{N}(\mu_\eta(1- \psi_\eta) + \psi_\eta \eta_{k-1}, \sigma_\eta)$ is an AR1 process. The marginal standard deviation of $\boldsymbol{\eta}$ is $\sigma_\eta = (\tau_\eta \kappa_\eta \sqrt{4\pi})^{-1}$, where $\tau_\eta$ and $\kappa_\eta$ are the scale and range parameters from the SPDE approach of \citet{lin11}, which is used to approximate the Mat\`ern field. The range parameter $\kappa_\eta$ determines the rate of spatial decorrelation (e.g., the distance for which correlations decline to 10\% is approximately $\sqrt{8}\kappa_\eta^{-1}$) and both $\kappa_\eta$ and $\tau_\eta$ scale the marginal sd.

\begin{table}[H]\centering
	\begin{tabular*}{\textwidth}{c p{11cm}}
		
		\multirow{1}{*}{ }
		\multirow{1}{*}{Parameter} & \multicolumn{1}{c}{\multirow{1}{*}{Description}}\\
		\midrule[1 pt]
		
		$\mu_\eta$ & Expected value of $\boldsymbol{\eta}$\\
		$\psi_\eta$ & Size autocorrelation parameter of $\boldsymbol{\eta}$ \\
		$\rho_\eta$ & Temporal autocorrelation parameter of $\boldsymbol{\eta}$ \\
		$\kappa_\eta$ & SPDE range parameter of $\boldsymbol{\eta}$ \\
		$\tau_\eta$ & SPDE scale parameter of $\boldsymbol{\eta}$ \\
		$\mu_\iota$ & Expected value of $\boldsymbol{\iota}$\\
		$\rho_\iota$ & Temporal autocorrelation parameter of $\boldsymbol{\iota}$ \\
		$\sigma_\iota$ & Marginal standard deviation of $\boldsymbol{\iota}$ \\
		$\alpha_{M c}$ & Expected rates of instantaneous natural mortality for each species $c \in \{1,...,v_C\}$\\
		
		\midrule[1 pt]
	\end{tabular*}
	\caption{Parameters that specify natural mortality.}
	\label{tab_M}
\end{table}

\subsubsection*{Fishing Mortality}
Harvesting is simulated on a grid with $v_F$ square cells that cover the domain of the fishery, $\Omega$. Each grid cell represents a potential fishing site and has area equal to the area fished by one unit of effort, $a_F$ (e.g., the area-swept by the tow of a dredge). The number of individuals with size $l \in Q_{c,k}$ at grid cell $f$ and time $t$ is interpolated from the values at the three closest triangulation nodes using barycentric coordinates, which calculate weightings of the vertex values to allot to any point within a triangle when interpolating the triangular lattice field into continuous space,
\begin{equation}
	n_{c,f,t,k} =  \frac{a_F}{a_S} \sum_{s \in \mathcal{N}_f} A_{f,s} n_{c,s,t,k} .
\end{equation}
Here, $A_{f,s}$ is the weight attributed to location $f$ from the node at location $s$ in the barycentric coordinate system, $a_S$ is the area associated with each of the triangulation nodes (i.e., the total area of $\Omega$ divided by the number of triangulation nodes within $\Omega$), and $\mathcal{N}_f$ is the neighborhood of location $f$.

The number of harvested individuals from the $i$th haul that belong to species $c$ with size $l \in Q_{c,k}$ is $x_{i,c,k} = n_{c,f,t,k} \zeta_{c,k}$. Here, $\zeta_{c,k}$ is the probability that an individual belonging to species $c$ with size $l \in Q_{c,k}$ is retained by the dredge. The harvest of these individuals is apportioned to the triangulation nodes $\in \mathcal{N}_f$ with proportion relative to $A_{f,s}$.

The total annual catch for species $c$ can be divided into catches representing scientific surveys and catches representing commercial harvesting. The number of survey sites and their distribution are user-specified. The distribution of commercial catch is determined by \autoref{alg_harvest} (\autoref{app}), which supports the specification of preferential targeting, area closures, and other site selection constraints (e.g., economic constraints can be incorporated through the specification of lower site selection probabilities for areas further away from ports).
%

The harvest algorithm has two distinct configurations that are specified using different combinations of the parameters $F^\text{int}$, $p^\text{targ}$, and $\mathbf{p}^\text{loc}$.
The first configuration simulates vessels systematically fishing across $\Omega$, with harvesting focused on areas of higher abundance. 
The level of fishing intensity across space is controlled by the parameters, $F^\text{int}$ and $p^\text{targ}$, where the former determines the size of the sampling region for a given time step and the latter is the probability of sampling from a site where the biomass is above the median. It is worth noting that this setup does not use the third parameter, $\mathbf{p}^\text{loc}$.
The second configuration uses the entire domain as the sampling region for each time step (i.e., $F^\text{int} \approx 0$) and specifies the site selection probabilities as the product of the preferential targeting probability (i.e., $p^\text{targ}$ for a site where the biomass is above the median and $1-p^\text{targ}$ otherwise), and $\mathbf{p}^\text{loc}$, which represents fixed site-selection probabilities specific to each cell in the harvesting grid. This specification allows users to account for economic considerations such as a sites distance from a port and other factors (e.g., depth, substrate type and other accessibility constraints) that may make areas more or less likely to be targeted.

\begin{figure}[H]	
	\vspace*{0in}		
	\centerline{\includegraphics[scale=1, trim = {0cm 0cm 0cm 0cm}]{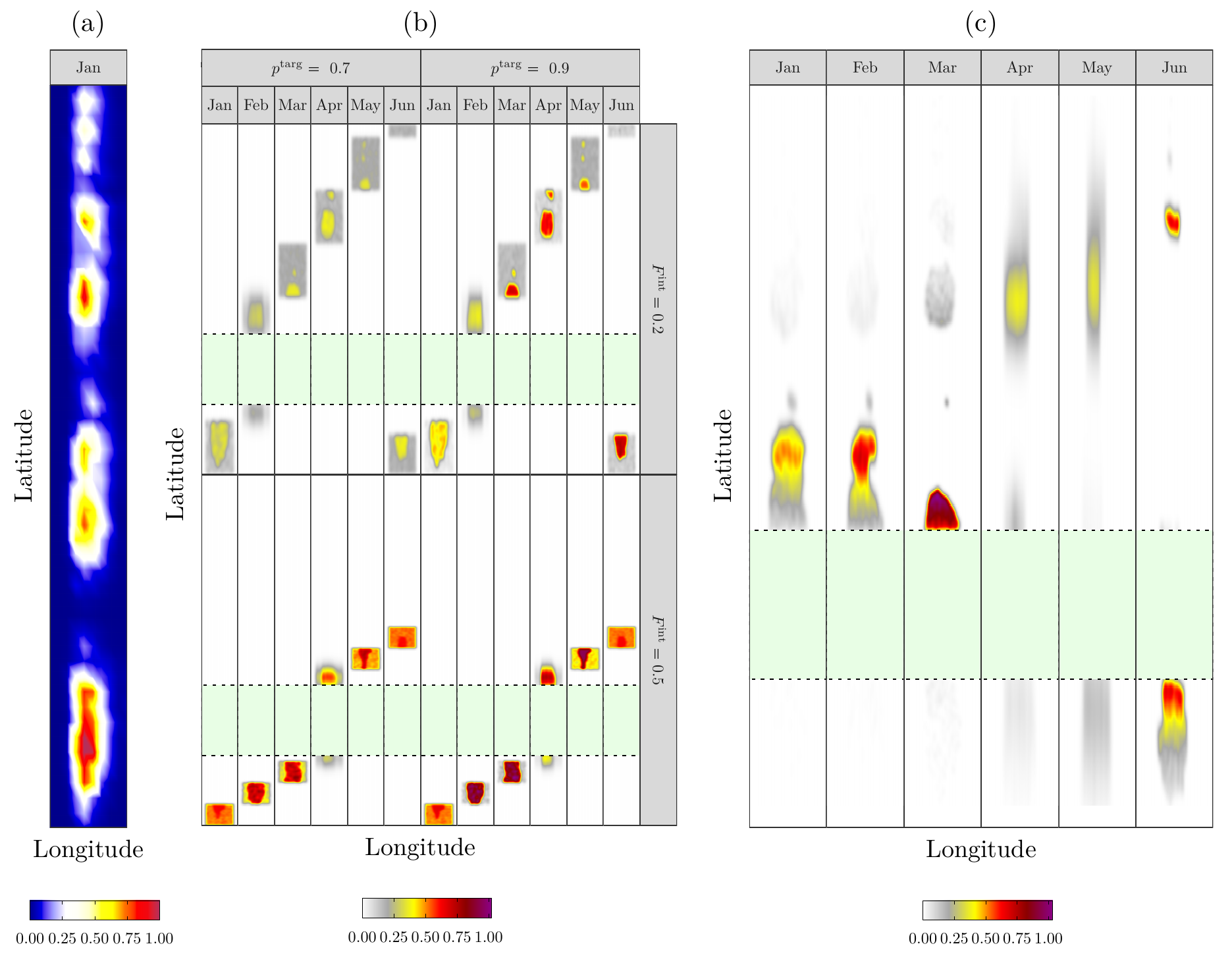}}
	\caption{Plot (a) shows a spatial density map of the distribution of pre-harvest biomass (kg m$^{-2}$) for a fishery simulated using the spatialSim OM, and plots (b) and (c) show spatial density maps of the distribution of fishing effort resulting from different configurations of the harvest algorithm (\autoref{alg_harvest}) when applied to this fishery. In the former, each element of $\mathbf{p}^\text{loc}$ was $1$ and harvesting was specified to systematically move across the domain with each time step for four scenarios that were specified using different values of $p^\text{targ}$ and $F^\text{int}$. Contrastingly, in plot (c) the probability of sampling commercial catch from a site given its geographic location, $p^\text{loc}_f$, was specified to exponentially decrease as the target location got further away from the domain's centroid and the other two parameters were specified as $p^\text{targ}=1$ and $F^\text{int}=0.1$. In all scenarios, the green shaded region within the dotted lines represents an area closed to commercial fishing.}
	\label{plot_fishing}
\end{figure}

%
%
An example of how the harvest algorithm can be specified to simulate different fishing patterns is presented in \autoref{plot_fishing}. In this example, commercial harvesting is simulated from January to June and the parameters $p^\text{targ}$, $F^\text{int}$, and $\mathbf{p}^\text{loc}$ are varied to illustrate their influence. In addition, to demonstrate how an area-closure could work, part of the domain (shaded in green) is excluded from the commercial sampling zone. A detailed overview of the user-set harvest parameters and functions is presented in \autoref{tab_F}.

\begin{table}[H]\centering
	\begin{tabular*}{\textwidth}{c p{11cm}}
		\multirow{1}{*}{ }
		\multirow{1}{*}{Parameter/} & \multicolumn{1}{c}{\multirow{2}{*}{Description}}\\
		\multirow{1}{*}{Function} & \\
		\hline
		$\mathcal{F}^\zeta(l; \boldsymbol{\theta}_c)$ & Function that determines the probability of an individual with size $l$ being retained by fishers given parameters $\boldsymbol{\theta}_c$, for each species $c \in \{ 1,...,v_C \}$\\
		$F^\text{lim}_c$ & Annual harvest limit for each species $c \in \{ 1,...,v_C \}$\\
		$F^\text{int}$ & Parameter controlling intensity of commercial harvesting\\	
		$p^\text{targ}$ & Probability of sampling commercial catch from a site with biomass above the median\\		
		$\mathbf{p}^\text{loc}$ & Vector containing the probabilities of sampling commercial catch from each grid cell given their geographic locations\\	
		$f^\text{one}_c$ & Site to begin commercial harvesting for each species $c \in \{ 1,...,v_C \}$\\
		$\mathbf{f}^\text{targ}_{c}$ & Vector containing indices to grid cells that make up the commercial target zone within $\Omega$ for each species $c \in \{ 1,...,v_C \}$\\
		$\mathbf{f}^\text{surv}_{c,t}$ & Vector containing indices to grid cells that will be sampled for scientific surveys for each species $c \in \{1,...,v_C\}$ and time $t \in \{1,..,n_T\}$\\
		\hline
	\end{tabular*}
	\caption{Parameters and functions that specify harvesting dynamics.}
	\label{tab_F}
\end{table}

\subsubsection*{Recruitment}
The number of species $c$ individuals that are recruited at location $s$ between time $t$ and $t+1$ with size $l \in Q_{c,k}$ depends on: a scaled Beverton-Holt stock-recruit function $\sr(\cdot)$ \citep{bev57} of the biomass of species $c$ individuals that are sexually mature at time $t$ averaged across the nodes within $r^\text{range}_c$ km of node $s$, giving the overall term $r_{0 c}\mathrm{SR}( \widebar{\ssb}_{c,s,t} )$; the proportion of recruitment allotted to size interval $Q_{c,k}$ given $\Delta t$, $\upsilon_k$; the probability of a species $c$ individual recruiting conditional on the environmental conditions at location $s$ between time $t$ and $t+1$, $\xi_{c,s,t}$; a recruitment cutoff threshold for species $c$, $r^\text{cut}_c$; and random spatiotemporal variability $\epsilon_{c,s,t}$,
\begin{equation}
	\label{eq_omr}
	r_{c,s,t,k} = \begin{cases}
		r_{0 c}\mathrm{SR}( \widebar{\ssb}_{c,s,t}; \ssbnaught_{c}, h_c)
		\xi_{c,s,t}
		\mathrm{logistic}(\epsilon_{c,s,t}) \upsilon_k, & \frac{\widebar{\ssb}_{c,s,t}}{\ssbnaught_{c}} \ge r^\text{cut}_c \\
		0, & \frac{\widebar{\ssb}_{c,s,t}}{\ssbnaught_{c}} < r^\text{cut}_c.
	\end{cases}
\end{equation}

\begin{figure}[H]	
	\vspace*{0in}		
	\centerline{\includegraphics[scale=1, trim = {0cm 0cm 0cm 0cm}]{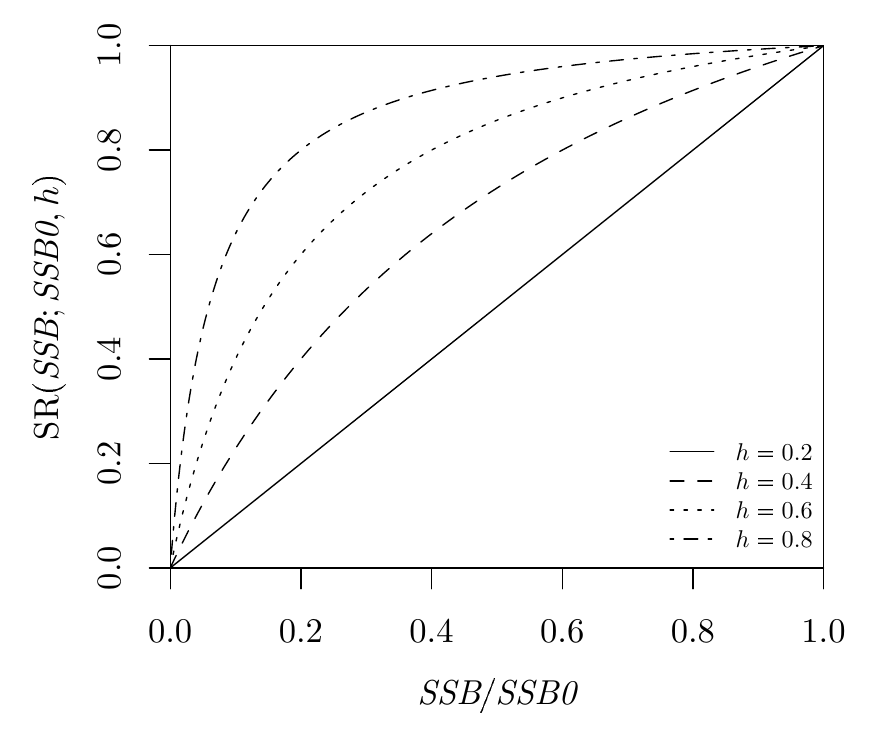}}
	\caption{The effect of the steepness parameter, $h$, on the Beverton-Holt stock-recruit function.}
	\label{plot_steepness}
\end{figure}

The stock-recruit function $\sr(\cdot)$ is parameterized in terms of the steepness parameter, $h_c$, which is the ratio of recruitment when $\widebar{\ssb}_{c,s,t}$ is 20\% of $\ssbnaught_{c}$ to recruitment when $\widebar{\ssb}_{c,s,t}$ is equal $\ssbnaught_{c}$ \citep{fra92} (\autoref{plot_steepness}), where $\ssbnaught_{c}$ is the expected unfished biomass of individuals belonging to species $c$ that are sexually mature at a location with above average environmental conditions for recruitment (see the section describing initialization for further details).
The biomass of species $c$ individuals that are sexually mature at time $t$ averaged across the spatial nodes that are within $r^\text{range}_c$ km of node $s$ is mathematically defined as,
\begin{equation}
	\widebar{\ssb}_{c,s,t} = \frac{\sum_{s^*} I_{s,s^*} SSB_{c,s^*,t} }{ \sum_{s^*}I_{s,s^*}}.
\end{equation}
where $I_{s,s^*}$ is an indicator variable that is equal to one if location $s^*$ is within $r^\text{range}_c$ km of location $s$ and zero otherwise, and $\textit{SSB}_{c,s,t} = \sum_k (n_{c,s,t,k} - x_{c,s,t,k})\bar{m}_{c,k} \bar{w}_{c,k}$ is the total biomass of species $c$ individuals that are sexually mature at location $s$ and time $t$ after accounting for harvesting.
Here, $\bar{m}_{c,k}$ is mean proportion of mature species $c$ individuals with size $l \in Q_{c,k}$ and $\bar{w}_{c,k}$ is the expected weight of a species $c$ individual with size $l \in Q_{c,k}$ (\autoref{tab_growth}).

The random spatiotemporal variability, $\boldsymbol{\epsilon}$, is a Gaussian stochastic process with correlation between species,  time, and space, $\mathrm{vec}(\boldsymbol{\epsilon}) \sim \mathrm{MVN}(\mathbf{1}\mu_\epsilon, \sigma_\epsilon \mathbf{C}_\epsilon \otimes \mathbf{T}_\epsilon \otimes \mathbf{S}_\epsilon)$. Here, $\mu_\epsilon$, is the marginal mean of $\boldsymbol{\epsilon}$, $\sigma_\epsilon$ is the marginal sd of $\boldsymbol{\epsilon}$, $\mathbf{C}_\epsilon$ is a user-defined species correlation matrix, and $\mathbf{T}_\epsilon$ and $\mathbf{S}_\epsilon$ are correlation matrices representing the temporal and spatial dimensions, respectively. If the temporal and species dimensions are marginalized, $\boldsymbol{\epsilon} \sim \mathrm{MVN}(\mathbf{1}\mu_\epsilon, \sigma_\epsilon \mathbf{S}_\epsilon)$ is a Mat\`ern field and if the species and spatial dimensions are marginalized, $\epsilon_t \sim \mathrm{N}(\mu_\epsilon(1- \rho_\epsilon) + \rho_\epsilon \epsilon_{t-1}, \sigma_\epsilon)$ is an AR1 process. The Mat\`ern field is approximated using the same approach that is used for the spatial field on natural mortality. Hence, the marginal standard deviation of $\boldsymbol{\epsilon}$ is $\sigma_\epsilon = (\tau_\epsilon \kappa_\epsilon \sqrt{4\pi})^{-1}$, where $\kappa_\epsilon$ and $\tau_\epsilon$ are the scale and range parameters from the SPDE approach.

\begin{figure}[H]	
	\vspace*{0in}		
	\centerline{\includegraphics[scale=1, trim = {0cm 0cm 0cm 0cm}]{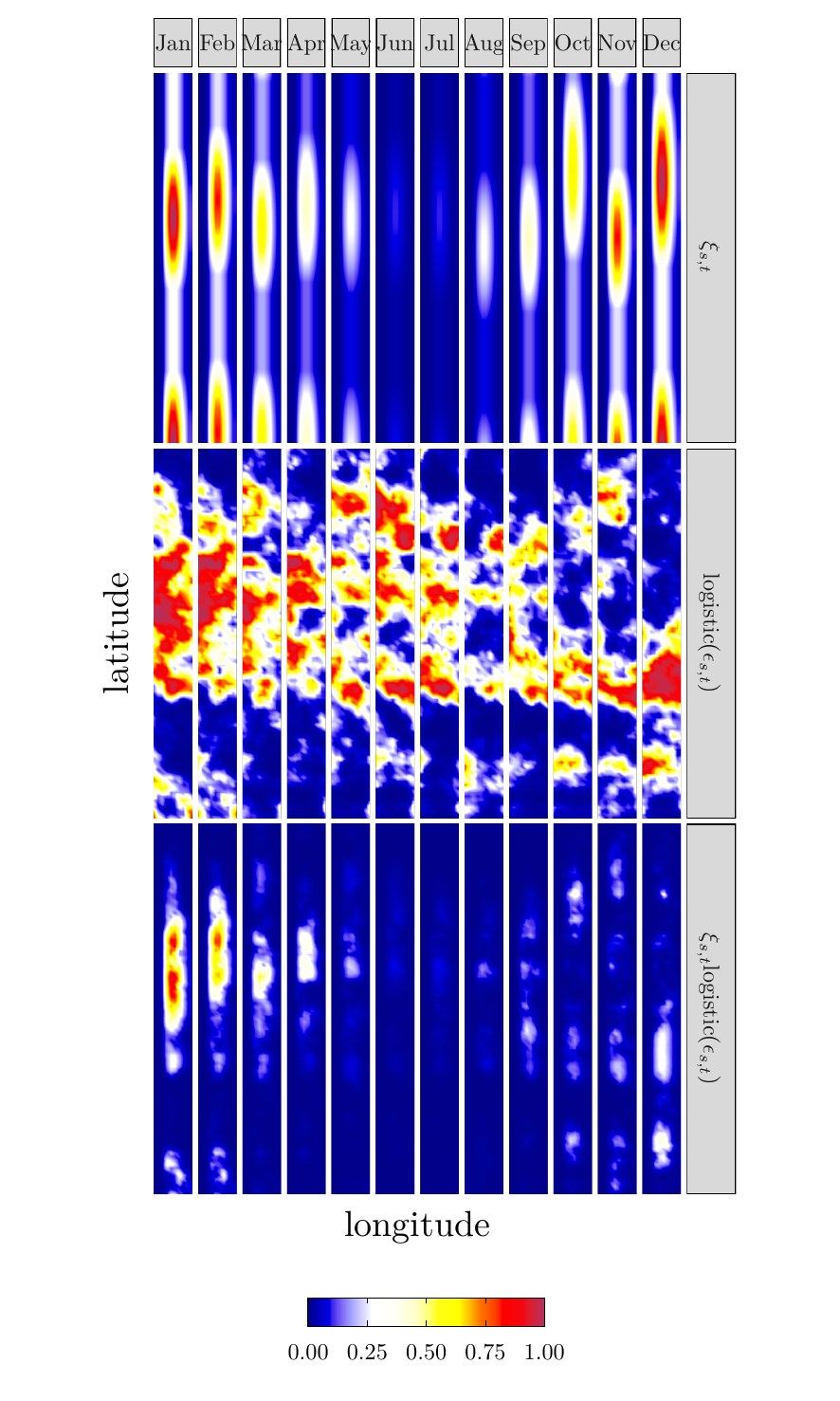}}
	\caption{An example of how $\boldsymbol{\xi}$ and $\boldsymbol{\epsilon}$ can be used to specify spatiotemporal variation in the recruitment of individuals of a given species.}
	\label{plot_recruit}
\end{figure}

The proportion of new recruits allotted to each size interval and the species specific probabilities of individuals recruiting conditional on the environmental conditions between each time period and at each location are defined externally by the user. An example of how $\boldsymbol{\xi}$ and $\boldsymbol{\epsilon}$ can be used to specify spatiotemporal variation in the recruitment of individuals is presented in \autoref{plot_recruit}. In this example, $\boldsymbol{\xi}$ includes a seasonal scaling effect and randomly shifting depth-zones that are optimal for recruitment. A detailed overview the parameters and functions that are used to specify recruitment is presented in \autoref{tab_R}.

\begin{table}[H]\centering
	\begin{tabular*}{\textwidth}{c p{11cm}}
		
		\multirow{1}{*}{ }
		\multirow{1}{*}{Parameter/} & \multicolumn{1}{c}{\multirow{2}{*}{Description}}\\
		\multirow{1}{*}{Function} & \\
		\midrule[1 pt]
		
		$\mathcal{F}^m(l; \boldsymbol{\theta}_c)$ & Function determining the expected proportion of sexually mature individuals in the population given their size $l$ and parameters $\boldsymbol{\theta}_c$, for each species $c \in \{ 1,...,v_C \}$\\ 
		$r_{0c}$ & Parameter used to scale recruitment for each species $c \in \{ 1,...,v_C \}$ \\ 
		$h_c$ & Steepness parameter of the stock-recruit relationship for each species $c \in \{ 1,...,v_C \}$ \\
		$\upsilon_k$ & Proportion of recruits allotted to each size interval $k \in \{ 1,...,v_L \}$ \\
		$\xi_{c,s,t}$ & Recruitment probability specific to each species $c \in \{1,...,v_C\}$ conditional on the environmental conditions between each time point $t \in \{1,...,v_T\}$ at each location $s \in \{1,...,v_S\}$\\	
		$r^\text{range}_c$ & Spawning range of sexually mature individuals for each species $c \in \{ 1,...,v_C \}$ \\
		$r^\text{cut}_c$ & Recruitment threshold for each species $c \in \{ 1,...,v_C \}$ \\
		$\mathbf{C}_\epsilon$ & Species correlation matrix of $\boldsymbol{\epsilon}$\\
		$\rho_{\epsilon}$ & Temporal autocorrelation parameter of $\boldsymbol{\epsilon}$ \\
		$\kappa_\epsilon$ & SPDE range parameter of $\boldsymbol{\epsilon}$ \\
		$\tau_\epsilon$ & SPDE scale parameter of $\boldsymbol{\epsilon}$ \\
		
		\midrule[1 pt]
	\end{tabular*}
	\caption{Parameters and functions that specify recruitment.}
	\label{tab_R}
\end{table}

\subsection*{Initialization}
\label{sec:init}
The state of the population at time $t = 0$ (i.e., the beginning of the first time period in the first year of harvesting) is calculated by executing the population dynamics (\ref{eq_N}) for $v_B$ prior time steps that act as an initialization period. At the start of the first time interval in the initialization period, the number of individuals belonging to each species is equal to zero. During the first two thirds of the $v_B$ steps, the SR relationship is equal to 1 for all locations and species: that is, $\widebar{\ssb}_{c,s,b}=\ssbnaught_{c}$ for $c \in \{ 1,....,v_C \}$, $s \in \{1,...,v_S\}$, and $b \in \{1,...,2v_B/3\}$. The expected unfished biomass of individuals belonging to species $c$ that are sexually mature at a location with above average conditions for recruitment is calculated from the biomass values obtained during the second third of the initialization period,
\begin{equation}
	\label{eq_ssb0}
	\ssbnaught_{c} = \frac{3\sum_b\sum_s \widebar{\ssb}_{c,s,b}}{v_{S^\text{opt}}}, \text{ ~ } \frac{v_B}{3} < b \le \frac{2v_B}{3} ,
\end{equation}
where $v_{S^\text{opt}}$ is the total number of spatial nodes where $\xi_{c,s,b} \ge \sum_s \sum_b \xi_{c,s,b} v_S^{-1} v_B^{-1}$, for $b \in \{v_B/3+1,...,  2v_B/3\}$. During the final third of the $v_B$ steps, recruitment is calculated using (\ref{eq_omr}). This period provides time for the number of individuals at the sub-optimal recruitment locations to converge to lower values. To ensure $\ssbnaught_{c}$ is calculated over a number of full recruitment cycles, $v_B/v_P$ must be an integer divisible by 3, where $v_P$ is the number of time periods that make up an annual cycle.

\section*{Case study: New Zealand surfclams}

\subsection*{Model parameters}
In New Zealand, the total extent of suitable surfclam habitat exceeds $2300$ km$^2$ and densities are typically in excess of $200$ tonnes per km$^2$ \citep{whi15}, suggesting that a population biomass of $500\,000$ t or more is possible. However, total annual catch has not yet reached $800$ t. To showcase some of the model's features and demonstrate execution time, a scenario was specified representing a hypothetical fishery applied to two New Zealand surfclam species, \textit{Spisula aequilatera} and \textit{Mactra murchisoni}. Simulations were carried out for three scenarios, which differed in their specified level of commercial fishing intensity across space. The OM scenarios are denoted here as OM$_\text{high}$, OM$_\text{med}$, and OM$_\text{low}$, where the subscript represents the associated level of fishing intensity. Each scenario was replicated 100 times and the specified parameters and functions were primarily based on published data for the Manawatu region. A detailed overview is presented in \autoref{app:ompars}.

The spatial domain was $56$ km$^2$, which is around the same size as the fishable area of the Manawatu fishery that the model's parameters were based on. It comprised $500$ triangulation nodes, $557\,433$ harvest cells and was divided into seven latitudinally extending bathymetrical depth bands ranging from 1 to 7 meters deep, with each representing a one meter change in depth. The fishery was simulated for a period of 15 years using monthly time increments. The first five years represented a period of unsustainable fishing and the ten years that followed were used to demonstrate the rate of recovery given no harvesting.

The growth of individuals from each species followed a von Bertalanffy curve \citep{ber38} with the parameterization described by \citet{fab65}. The von Bertalanffy parameters were set to the average values estimated by \citet{cra93}, \citet{cra96}, and \citet{cra01a}. The sd of the growth increment on the log scale was set to $0.15$ for each species and the size distribution of each population was represented using 10 size intervals.

Harvesting was simulated on a grid where each cell represented a single dredge tow corresponding to an area-swept of approximately $\text{100 m}^2$. Catch limits were set at 1000 t per annum and scientific biomass surveys were simulated at 50 random sites uniformly distributed across $\Omega$ per year. The commercial fishing target zones were defined as the entirety of $\Omega$ for both species. Selectivity was 0.95 for individuals in the first size-class and 1 otherwise. The probability of a commercial vessel harvesting from a cell with biomass above the median was set to $p^\text{targ}=\ptarg$, which corresponds to fishers having very good knowledge about the distribution of individuals in the areas of harvesting.

\begin{figure}[H]	
	\vspace*{0in}		
	\centerline{\includegraphics[scale=1, trim = {0cm 0cm 0cm 0cm}]{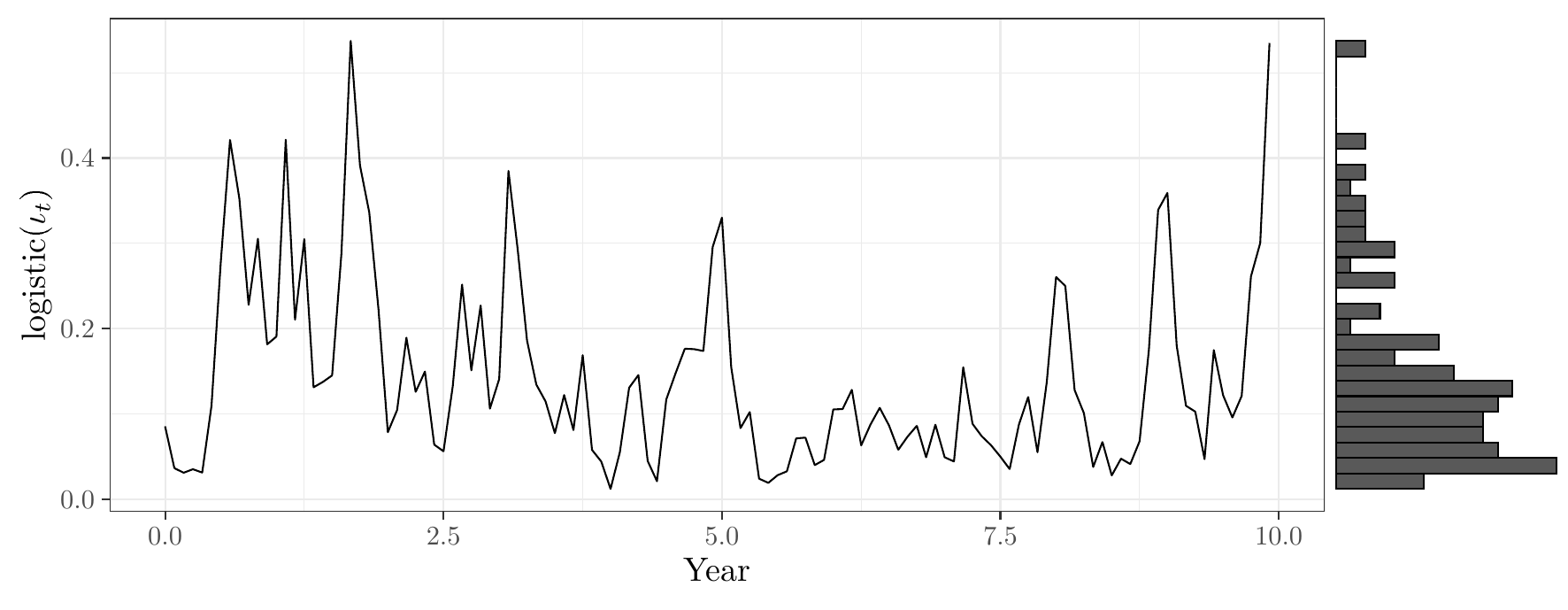}}
	\caption{The simulated logistic transformed $\boldsymbol{\iota}$ values for each time period in the case study with a marginal histogram.}
	\label{plot_I}
\end{figure}

\begin{figure}[H]	
	\vspace*{0in}		
	\centerline{\includegraphics[scale=1, trim = {0cm 0cm 0cm 0cm}]{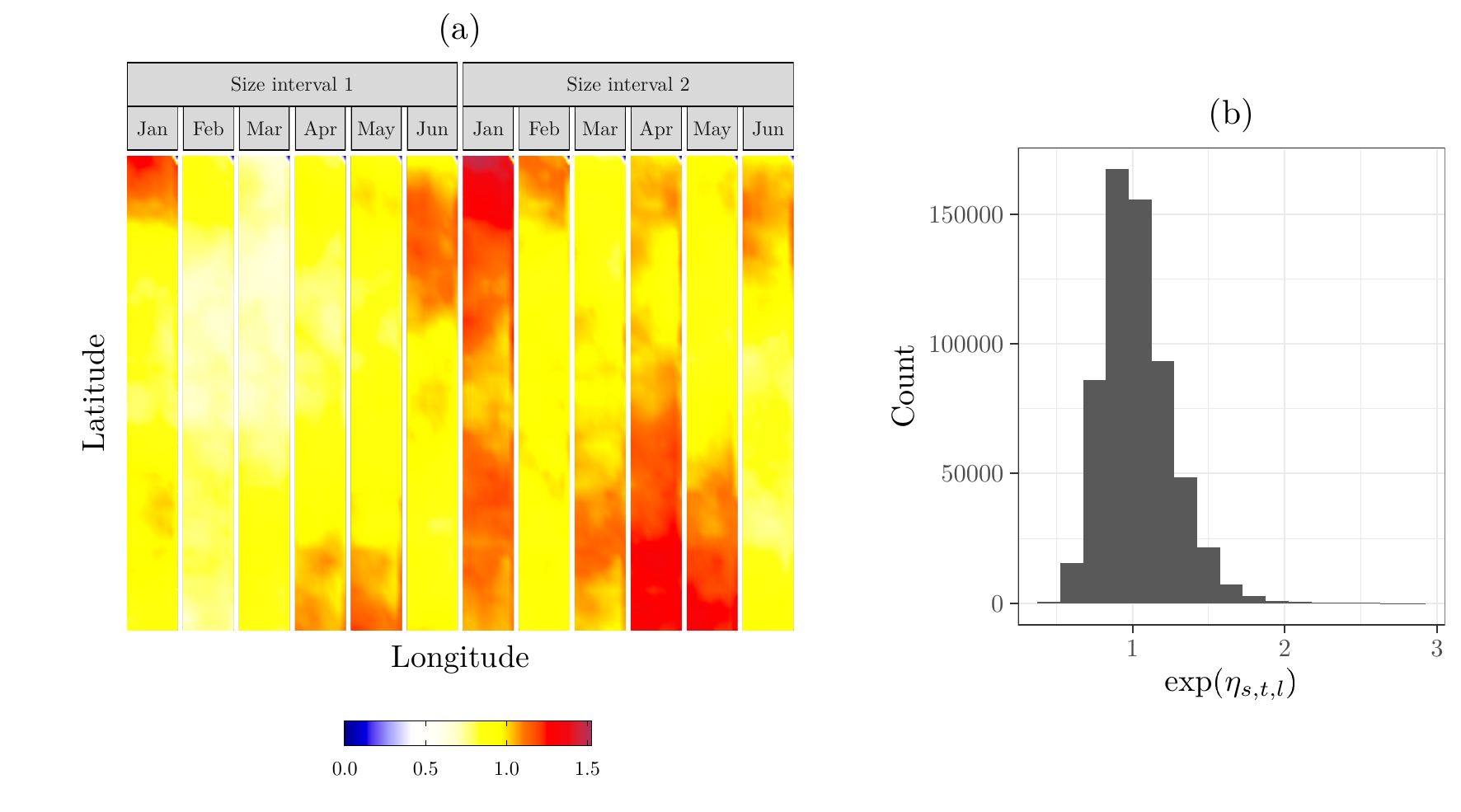}}
	\caption{The plot on the left (a) is a spatial density map of the simulated exponential transformed $\boldsymbol{\eta}$ values associated with the first two size intervals for the first 6 months of the first year of the case study, and the plot on the right (b) is a marginal histogram of all of the simulated exponential transformed $\boldsymbol{\eta}$ values.}
	\label{plot_H}
\end{figure}

The parameters of $\boldsymbol{\iota}$ were set to $\sigma_\iota=\sigmaIota$, $\rho_\iota=\rhoIota$, and $\mu_\iota=\muIota$. This combination of values produces a right skewed distribution with moderate variability and a high level of temporal autocorrelation, equating to relatively long periods of lower mortality rates followed by shorter periods with higher rates (\autoref{plot_I}). The autocorrelation parameters of $\boldsymbol{\eta}$ were set to $\psi_\eta=\psiEta$ and $\rho_\eta=\rhoEta$, which corresponds to strongly correlated mortality rates for individuals of similar size and strongly correlated mortality rates between time periods, given $\boldsymbol{\iota}$ (\autoref{plot_H}). The scale and range parameters of $\boldsymbol{\eta}$ were set to $\tau_\eta=\tauEta$ and $\kappa_\eta=\kappaEta$, which correspond to relatively small amounts of spatial variability in $\boldsymbol{\eta}$ with correlation decaying slowly over large distances (the distance for which correlation declines to 10\% was approximately 11.8 km). The expected rates of instantaneous natural mortality for individuals belonging to each species were set to $\boldsymbol{\alpha_M} =\alphaM$, which are the averaged estimates of \citet{cra93}.

The probability of a species $c$ individual recruiting conditional on the environmental conditions between time $t$ and $t+1$ at location $s$, was specified using two independent components. These included (1) an annual seasonality term consistent with research on the gametogenic cycles of surfclam species in New Zealand \citep{nuh13, not14} and (2) a time-invariant bathymetrical term based on the stratified biomass survey by \citet{whi12}.

\begin{figure}[H]	
	\vspace*{0in}		
	\centerline{\includegraphics[scale=1, trim = {0cm 0cm 0cm 0cm}]{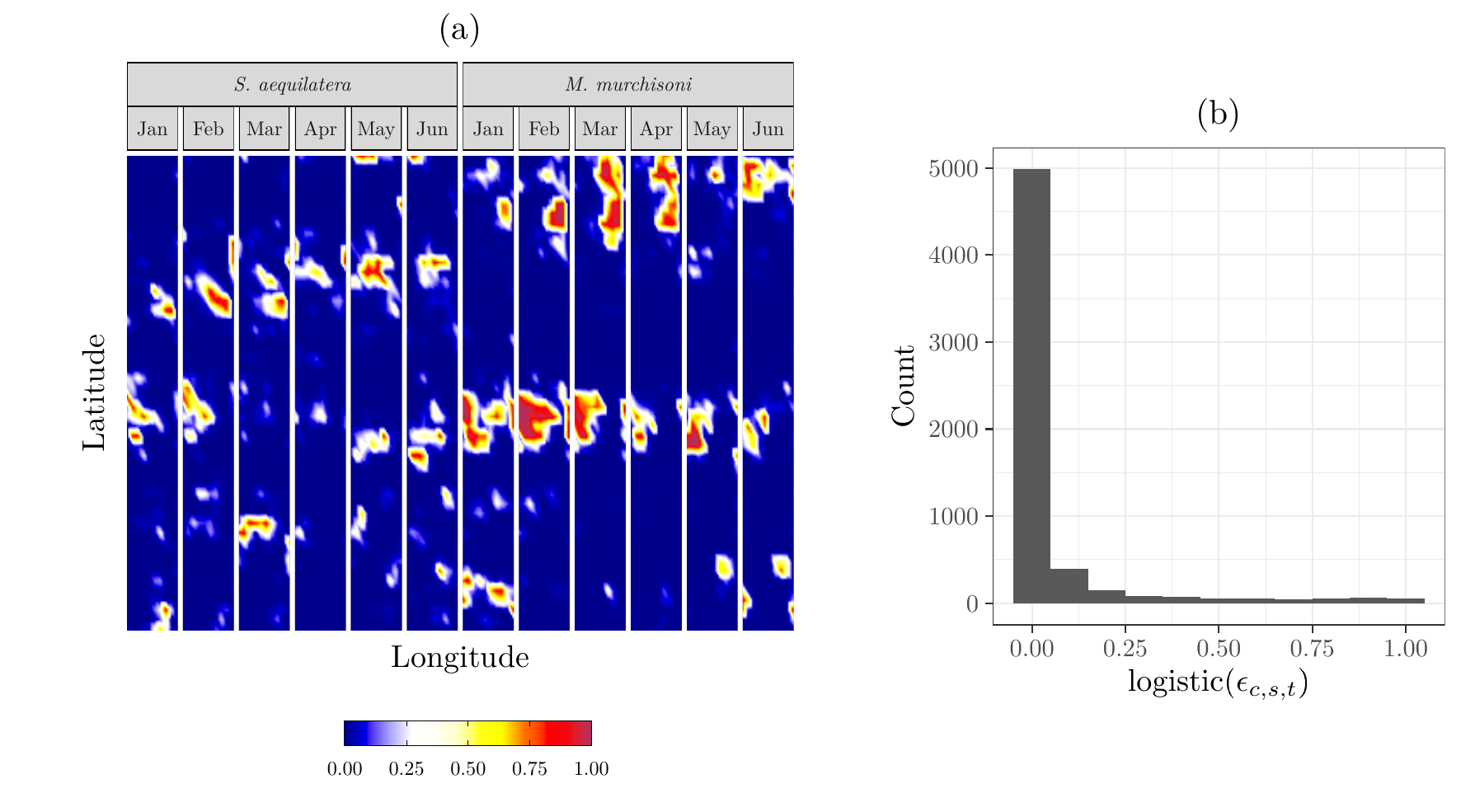}}
	\caption{The plot on the left (a) is a spatial density map of the simulated logistic transformed $\boldsymbol{\epsilon}$ values for the first 6 months of the first year of the case study and the plot on the right (b) is a marginal histogram of all of the simulated logistic transformed $\boldsymbol{\epsilon}$ values.}
	\label{plot_E}
\end{figure}

The marginal mean of $\boldsymbol{\epsilon}$ was set to $\muEps$, which produced a very sparse and patchy recruitment pattern (\autoref{plot_E}) that resulted in distinct beds of individuals. The temporal autocorrelation parameter of $\boldsymbol{\epsilon}$ was set to $\rho_\epsilon=\rhoEps$, which corresponds to a high level of temporal autocorrelation in recruitment rates. The scale and range parameters of $\boldsymbol{\epsilon}$ were set to $\tau_\epsilon=\tauEps$ and $\kappa_\epsilon=\kappaEps$, which correspond to relatively large amounts of spatial variability in $\boldsymbol{\epsilon}$ with correlation decaying quickly over relatively short distances (the distance for which correlations decline to 10\% is approximately 1.6 km). The species correlation matrix of $\boldsymbol{\epsilon}$ was set as the identity matrix. This combination of parameters produces a patchiness in the spatial distribution of the populations that is consistent with what was reported by \citet{whi12}.

The spawning range and steepness parameters of the stock-recruit function were set to $r^\text{range}_c = 0.5$ km and $h_c=0.9$ for each species. These values were not based on any real data, but were chosen to support the simulation of serial depletion over the five year harvesting period. The values of $\boldsymbol{r_0}$ were chosen such that each species had biomass densities in the ranges indicated by \citet{had96}, \citet{tri08a}, \citet{whi12}, and \citet{whi15}.

To provide insight into the computational cost of running the model, the first five years of the OM$_\text{high}$ scenario was repeated on 20 different domains ranging in size from $56$ km$^2$ to $224$ km$^2$ and the median execution times were recorded from ten replicates. The number of triangulation nodes and harvest cells changed in proportion with the area of $\Omega$ and ranged from 500 to 2000 nodes and $557\,433$ to $2\,244\,792$ cells, respectively. Execution times were recorded from an R session linked to the multithreaded Intel Math Kernel Library \citep{int09} on a desktop computer with an Intel 7700K processor and 32 gigabytes of memory with a clock frequency of 2667 MHz. The recorded times excluded the overheads associated with the initial setup process, which included creating a list specifying the model settings and simulating the random processes. 

\begin{figure}[H]
	\vspace*{0in}		
	\centerline{\includegraphics[scale=1, trim = {0cm 0cm 0cm 0cm}]{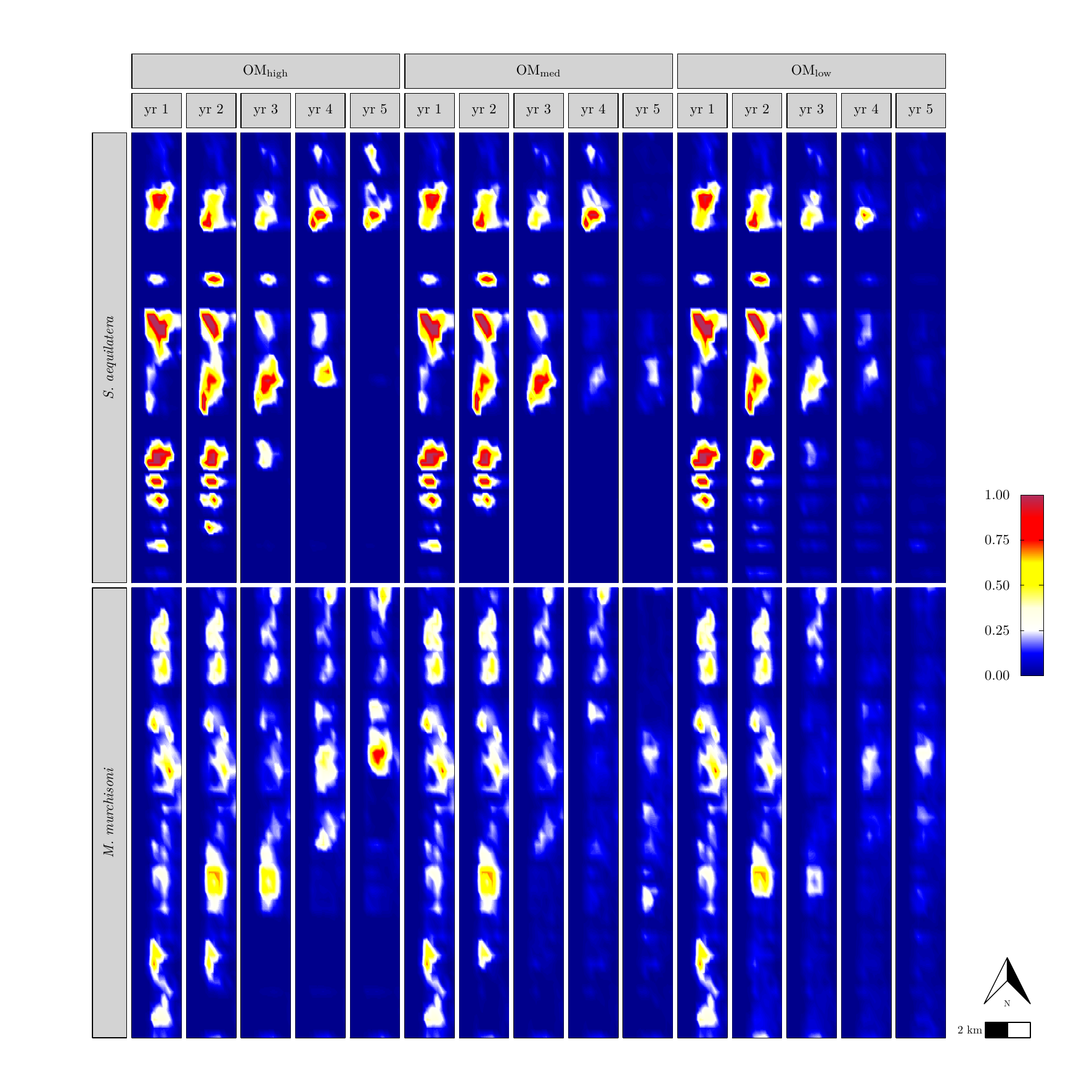}}
	\caption{The population densities (kg $\text{m}^{-2}$) of the simulated surfclam populations for one of the replicates of the three OM variants specifying high, medium, and low levels of fishing intensity across space.} 
	\label{biomass_map}
\end{figure}

\begin{figure}[H]	
	\vspace*{0in}		
	\centerline{\includegraphics[scale=1, trim = {0cm 0cm 0cm 0cm}]{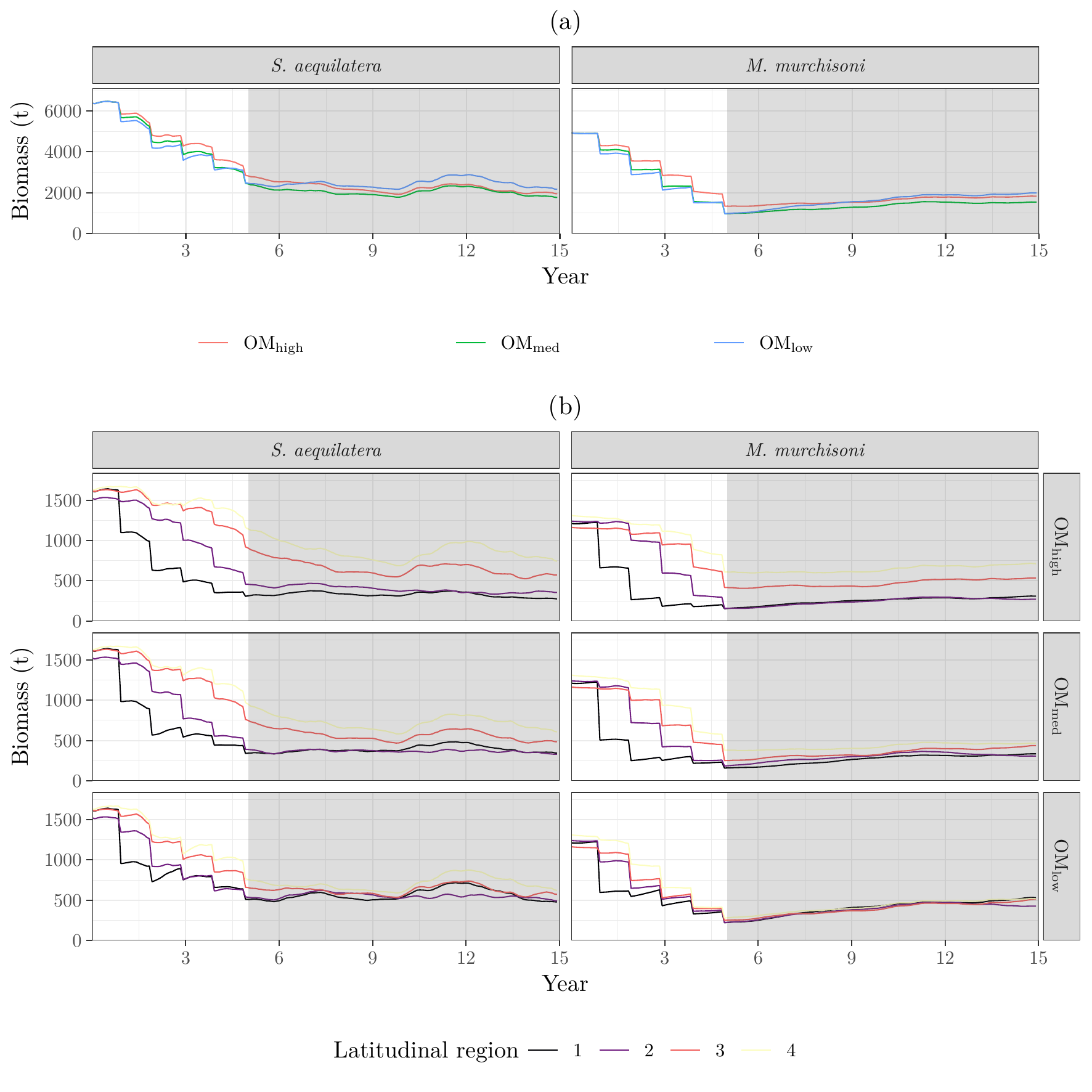}}
	\caption{Plot (a) shows the biomass of each simulated surfclam population across time for each of the three operating model scenarios averaged across the 100 replicates, and plot (b) shows the biomass of each simulated surfclam population aggregated into 4 equally sized latitudinal regions across time for each of the three operating model scenarios averaged across the 100 replicates. Latitudinal region 1 is the southern most followed by regions 2, 3, and 4. The shaded regions represent the time period with no harvesting.}
	\label{biomass_time}
\end{figure}

\begin{figure}[H]	
	\vspace*{0in}		
	\centerline{\includegraphics[scale=1, trim = {0cm 0cm 0cm 0cm}]{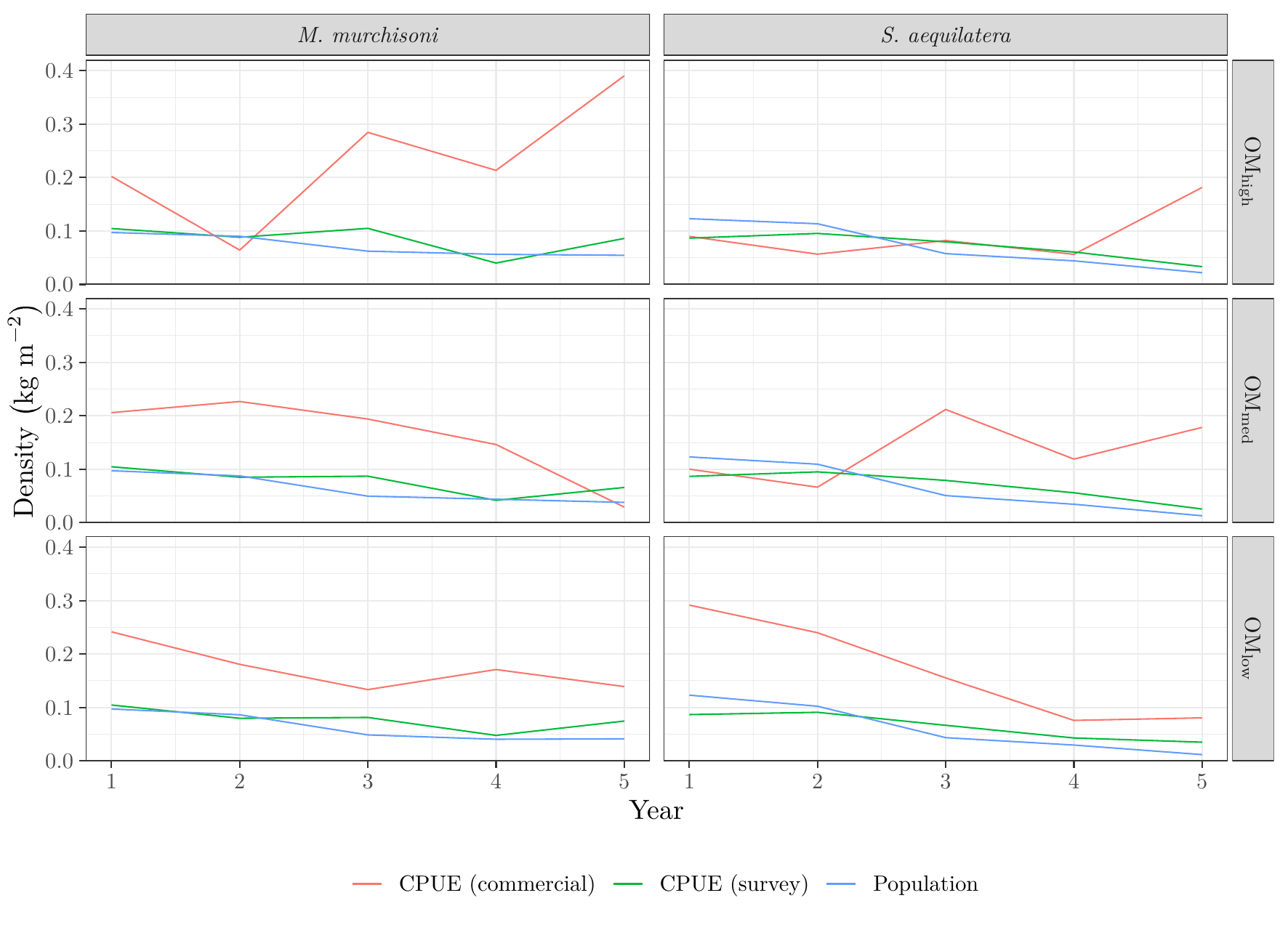}}
	\caption{The mean population densities, commercial CPUE, and survey CPUE across time for one of the replicates of the three OM variants specifying high, medium, and low levels of fishing intensity across space.}
	\label{plot_cpue}
\end{figure}

\begin{figure}[H]	
	\vspace*{0in}		
	\centerline{\includegraphics[scale=1, trim = {0cm 0cm 0cm 0cm}]{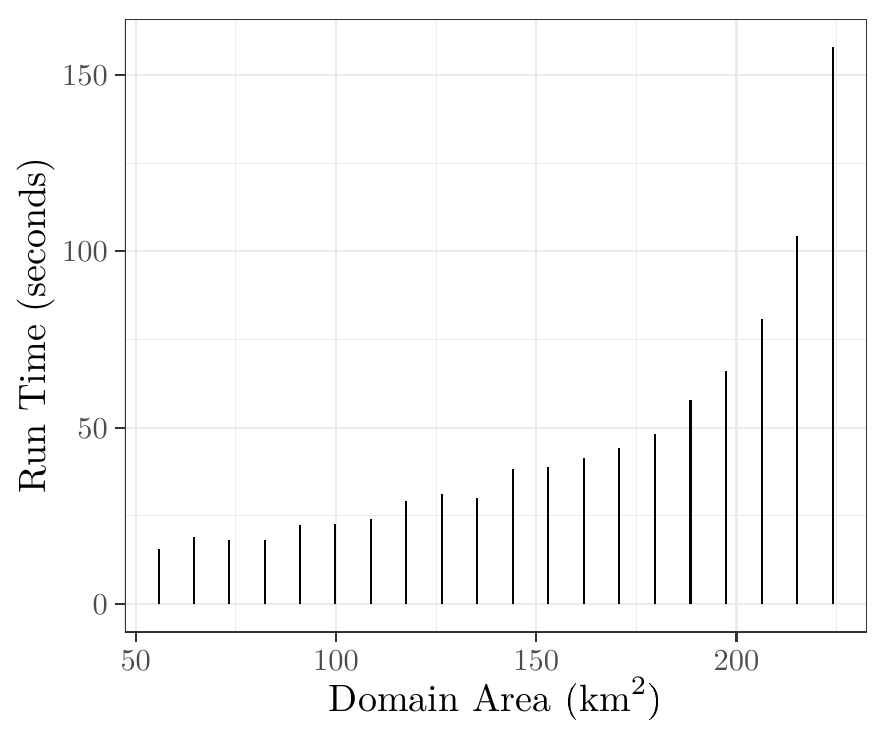}}
	\caption{Median execution times from 10 replicated sets of 20 model runs with different sized spatial domains that ranged from $56$ km$^2$ to $224$ km$^2$.}
	\label{plot_runtime}
\end{figure}

\subsection*{Results}
During the five year period of harvesting there were significant declines in the abundance of individuals and varying levels of spatial serial depletion among the different OM cases (\autoref{biomass_map}; \autoref{biomass_time}). During this period, the changes in abundance and spatial serial depletion were strongly correlated to the level of fishing intensity with the higher levels resulting in smaller decreases in population size and greater increases in the level of spatial serial depletion. At the highest level of fishing intensity, population levels in the southern half of the domain were almost completely wiped out, while population levels in the northern half of the domain were largely unaffected.

During the ten year period post-harvesting, the recovery rates in population abundance were limited for all scenarios. Overall, there was a slight increase in the abundance of \textit{M. murchisoni} individuals, but no substantial recovery in the simulated \textit{S. aequilatera} populations. Nevertheless, it is notable that the populations from the scenario with the lowest level of fishing intensity showed the strongest signs of a recovery, followed by the populations from the OM$_\text{high}$ and OM$_\text{med}$ scenarios. The reason for this is made clear in \autoref{biomass_time}, which shows the biomass trajectories across time aggregated into 4 equally sized latitudinal regions. The recovery rates associated with the OM$_\text{high}$ and OM$_\text{med}$ scenarios were primarily driven by increases in abundance in the northern parts of domain, which had seen less fishing. Alternatively, the recovery rates associated with the OM$_\text{low}$ scenario were driven by increases in abundance across the entire domain.




In all OM scenarios, temporal trends in stock depletion were captured reasonably well by the simulated survey CPUE data that represented 50 annual survey dredge tows per year (e.g., \autoref{plot_cpue}). In contrast, the simulated commercial CPUE data from the models with more intensive fishing regimes (OM$_\text{high}$ and OM$_\text{med}$) were often extremely poor indicators of the states of the fisheries (e.g., \autoref{plot_cpue}). This was because intensive harvesting was localized to one region each season. Consequently, in each subsequent fishing season harvesting was occurring in new areas with relatively high densities of individuals. In the case of OM$_\text{low}$, where harvesting was the least intensive, the simulated commercial CPUE was, on average, a reasonable measure of the biomass of each species.

Model execution times increased exponentially with domain size. The median execution times for each of the 20 different sized domains in the 10 replicated sets ranged between $15.6$ seconds for the case where the domain was $56$ km$^2$ to $158.0$ seconds for the case where the domain was $224$ km$^2$ (\autoref{plot_runtime}).

\section*{Discussion}
There are a wide range of different spatial OMs that have been described in the literature. These range from very simplistic models that partition a population across a small number of discrete zones \citep[e.g.,][]{pun09, fay11, ive13}, to extremely complex ecosystem models \citep[e.g.,][]{chi05, gra06, ful11} that, in practice, can only be run at low spatial resolutions due to computational constraints. Some authors have rightly inferred that rather than providing insight, the latter approach can lead to two things that are not understood, the real world system that was never properly understood to begin with, and a highly detailed model of it \citep{pao11, nee15}.

This has prompted recent attempts to develop models that fall between the overly simplistic and overly complex ecosystem models. An example of this is the Honeycomb model \citep{nee15}, which incorporates discrete spatial variability and fleet dynamics, but deliberately avoids incorporating superfluous elements that make understanding the model difficult. The interpretable simplicity of these kinds of model provide a much more user friendly experience than the complex ecosystem alternatives. Parameter setting is straightforward and the computational benefits associated with a simpler model enables users to model the system at much higher resolutions.

The model presented in this paper was developed with the same philosophy of interpretable simplicity that drove the development of the Honeycomb model. However, it is unique in that it makes it computationally feasible to simulate a fishery system at the same spatial resolution that it would operate in real world situations (e.g., the average area-swept by the tow of dredge). This efficiency is achieved by combining a spatially continuous GMRF model of the population dynamics with a harvest algorithm based on the highly efficient sampling routine described by \citet{efr06} that operates on an areal grid system containing the projected numbers-at-size.

The presented model was designed to help answer management questions for sedentary species, while allowing users to consider the implications of spatial heterogeneity, fleet dynamics, and spawning range. These are of fundamental importance to providing management advice that protects sedentary species against serial depletion. This was demonstrated in the case study, which showcased the simulation of three scenarios that each represented a fictional fishery comprising two species. The number of recruits at a given location was specified to depend on the average density of sexually mature individuals within a 500 m range of that location and the effect of three different levels of fishing intensity across space were compared.

The results demonstrated, that for the given scenarios, more intensive fishing across a spatial area can lead to lower levels of stock depletion in the short term, but in the longer term can potentially lead to higher levels of depletion due to reduced recruitment and slower recover rates. Another notable finding was that the commercial CPUE data generated from the two scenarios with more intensive fishing were completely uncorrelated with the abundance of individuals in the populations. In the scenario where harvesting was the most spatially extensive, the simulated CPUE data appeared to be a reasonable proxy for estimating total abundance. This phenomenon resulted from the more intensive fishing regimes producing highly localized commercial catch data that were generated from areas that had not been fished in previous seasons. In addition to these findings, the case study also demonstrated how the operating model can be used to generate CPUE data representing scientific surveys. In all of the presented cases, samples from $50$ random sites uniformly distributed across the fishery appeared to be a reasonably good proxy for estimating abundance.

The phenomenon that was observed in the case study, where the commercial CPUE data were uncorrelated with the trends in population abundance, is quite likely to occur in many of the worlds fisheries. Nevertheless, few existing stock assessment models are designed to model a fishery with data containing these features. In the fisheries literature there are three approaches that have been described to account for spatial effects \citep{pun19b}. These include (1) pre-processing data to remove spatial effects (e.g., CPUE standardization), (2) approximating a spatial distribution by size and/or age using selectivity \citep[e.g.,][]{wat14}, and (3) applying a model with spatially explicit population dynamics.

Recent research has primarily focused on categories (1) and (3), which has led to number of recent advancements, although, it is uncommon to see methods belonging to category (3) used in practice. In contrast, there has been a surge in the use of geostatistical CPUE standardization methods in stock assessments. This has been largely due to a geostatistical delta GLMM maximum likelihood estimator for abundance index standardization developed by \citet{tho15}, which has also been extended to handle multiple-species/categories \citep{tho19}. Some examples of recently developed models that incorporate spatially explicit population dynamics include a spatiotemporal size-structured model \citep{kri14}, a geostatistical delay-difference model \citep{tho15b}, and a geostatistical surplus production model \citep{tho17b}.

As technology develops it is becoming more economically feasible to record commercial catch data at the per-haul level. The OM presented in this paper could be a useful tool to aid in the development of new spatiotemporal assessment models designed to  leverage the additional information contained in these data (e.g., models that build on the work of \citealp{kri14}, \citealp{tho15b}, \citealp{tho17b}, and \citealp{tho19}) and, ultimately, to help quantify the costs and benefits of recording data in this way. In these applications, the spatialSim model's ability to simulate scenarios with realistic catch data (e.g., CPUE data with bias caused by localized depletion) could facilitate a more meaningful assessment than a lot the other OMs that are currently available.

The current version of the spatialSim model has a lot of potential for further development and its utility could be greatly increased by simply extending the range of options for specifying the growth, natural mortality, and recruitment processes. This is because the current specification of these processes are only appropriate for cases where it is reasonable to assume that growth is spatially homogeneous, natural mortality is proportionally related among species, and the populations self-recruit. 
Thus, the model could benefit from future work that adds support for spatially varying growth, different natural mortality correlation structures among species, and population connectivity (i.e., recruitment input from populations outside of the domain). 

There is also a very wide scope for extending the harvest algorithm. A couple of good options could include incorporating species interactions in the site selection process (e.g., bias towards areas with lower or higher diversities of the simulated species) and adding support for site selection biases based on the sizes of individuals (e.g., bias towards areas with larger individuals). These features would greatly enhance the multi-species aspect of the model and increase the range of management options that could be explored (e.g., maximizing catch while avoiding by-catch constraints or small-fish protocols).

The model presented here provides a way of carrying out a spatial MSE where data can be simulated to scale with the fishery. It adds more complexity and realism than a lot of the currently available alternatives, yet retains enough simplicity to allow the model and its parameters to be easily understood by users. This simplicity means the model is limited in the range of scenarios it can simulate and questions it can answer. Nevertheless, it was developed in a modular way that is easily extendable. Thus, this open-source project has a lot of potential for further development, which would help facilitate its use in a much wider range of simulation studies.

\section*{Acknowledgments}
I C. D. Nottingham thank the New Zealand Ministry for Primary Industries and the University of Auckland for the funding provided through the Ministry for Primary Industries Postgraduate Science Scholarship and the University of Auckland Doctoral Scholarship, which helped to make this project possible. We also thank R. Fewster and I. Tuck for their valuable input.

\newpage
\renewcommand{\bibname}{References} 
\bibliographystyle{apalike}
\bibliography{spatialSim_ms}

\newpage
\begin{appendices}
\label{appendices}

\counterwithin{figure}{section}
\counterwithin{table}{section}

\section{Algorithms} \label{app}

\newcommand\mycommfont[1]{\footnotesize\ttfamily\textcolor{gray}{#1}}
\SetCommentSty{mycommfont}

\SetKwProg{Fn}{Function}{}{}
\SetKwInOut{Input}{input}
\SetKwInOut{Output}{output}
\SetKwInOut{Vars}{variables}

The spatialSim harvest routine is presented in \autoref{alg_harvest}. Additionally, three sub-routines that are used within \autoref{alg_harvest} are presented. These include \autoref{alg_cumsum}, which returns the cumulative sums of a vector, \autoref{alg_ifelse}, which assigns each element of a vector as one of two values depending on whether the element of boolean input vector is TRUE or FALSE, and \autoref{alg_sample}, which takes a sample of size $n$ without replacement from the elements of an input vector based on specified selection probabilities for each element of the input vector. 

\renewcommand{\thealgocf}{A.\arabic{algocf}}

\begin{algorithm}
	$n = $ the size of the vector $\mathbf{x}$\\
	$y_i = \sum_{k=1}^i x_k$ for all $i \in \{ 1,...,n \}$\\
	\Return{$\mathbf{y}$}

	\caption{$\mathrm{cumsum}(\mathbf{x})$} 
	\label{alg_cumsum}
\end{algorithm}

\begin{algorithm}
	$n = $ the size of the vector $\mathbf{x}$\\
	\For{$i=1$ \textup{to} $n$}{
		\eIf{$x_i  =$ \textup{TRUE}}{$y_i = a_i$}{$y_i = b_i$}
	}
	\Return{$\mathbf{y}$}

	\caption{$\mathrm{ifelse}(\mathbf{x}, \mathbf{a}, \mathbf{b})$} 
	\label{alg_ifelse}
\end{algorithm}

\begin{algorithm}
	\Input{$\mathbf{x}$: Vector of size $m$ from which samples will be drawn\newline
	$n$: Number of samples to draw\newline
	$\mathbf{p}$: Vector containing the probabilities of drawing $x_i$ for $i \in \{1,...,m\}$
	}
	$r_i = $ simulate exponential(1)$/p_i$ for all $i \in \{ 1,...,m \}$\\
	$\mathbf{y} = $ $\mathbf{x}$ sorted by $\mathbf{r}$ in ascending order\\
	\Return{\textup{the the first} $n$ \textup{elements of} $\mathbf{y}$}
	\caption{sample$( \mathbf{x}, n, \mathbf{p} )$ \citep{efr06}.} 
	\label{alg_sample}
\end{algorithm}

\begin{algorithm}
	\Input{
	$b_f$: The selected biomass of individuals of a given species in each grid cell $f \in \{ 1,...,v_F \}$\newline
	$n_{f,k}$: The selected number of individuals of a given species in each grid cell $f \in \{ 1,...,v_F \}$ with size in each interval $k \in \{ 1,...,v_L \}$\newline
	$F^\textup{lim}$: Harvest limit (biomass)\newline
	$f^\textup{one}$: Grid cell to begin harvesting\newline
	$\mathbf{f}^\textup{targ}$: Vector containing indices to fishing grid cells that are in the area targeted by commercial fishers of a given species\newline
	$\mathbf{f}^\textup{surv}$: Vector containing indices to grid cells to be sampled for scientific surveys\newline
	$F^\textup{int}$: Parameter controlling the spatial intensity of commercial harvesting \newline
	$p^\textup{targ}$: Probability of fisher harvesting from a site with higher than the median biomass\newline
	$\mathbf{p}^\textup{loc}$: Vector containing the probabilities of sampling commercial catch from each grid cell given their geographic locations
	}

	$\mathbf{f}^\textup{targ} = $ elements of $ \mathbf{f}^\textup{targ} \not\in \mathbf{f}^\textup{surv}$\\
	$n^\textup{targ} = $ the size of the vector $\mathbf{f}^\textup{targ}$\\
	$f^\textup{one} = $ the index of the element of $\mathbf{f}^\textup{targ}$ closest to the cell $f^\textup{one}$\\
	$\mu = \mathrm{median}(\mathbf{b})$\\
	$n^\textup{samp} = \min\left(\ceil*{F^\textup{lim} / \mu}, \ceil*{0.9 n^\textup{targ}}\right)$\\
	$n^\textup{site} = \min\left( \ceil*{n^\textup{samp} / F^\textup{int}}, n^\textup{targ} \right)$\\
	
	\eIf {$n^\textup{site} < n^\textup{targ} - f^{\textup{one}}$} {
		$\mathbf{f}^\textup{site} = $ elements of $\mathbf{f}^\textup{targ}$ with indices 
		$\in \{ f^\textup{one},...,f^\textup{one} + n^\textup{site} \}$\\ 
		$f^\textup{one} = f^\textup{targ}_{f^\textup{one} + n^\textup{site} + 1}$\\	
	} {
		$n^\textup{site1} = n^\textup{targ} - f^\textup{one}$\\
		$n^\textup{site2} = n^\textup{site} - n^\textup{site1}$\\
		$\mathbf{f}^\textup{site} = $ elements of $\mathbf{f}^\textup{targ}$ with indices
		$\in \{1,...,n^\textup{site2}, f^\textup{one},...,n^\textup{targ} \}$\\
		$f^\textup{one} = f^\textup{targ}_{n^\textup{site2} + 1}$
	}

	$\mathbf{b}^\textup{site} = $ elements of $\mathbf{b}$ with indices $\in \mathbf{f}^\textup{site}$\\
	$\mathbf{p}^\textup{site} = \mathrm{ifelse}(\mathbf{b}^\textup{site} > \mu,  p^\textup{targ} \mathbf{p}^\textup{loc}, (1 -  p^\textup{targ})\mathbf{p}^\textup{loc})$

	$\mathbf{x}^\textup{bio} = $ vector of size $n^\textup{samp}$ with elements equal to zero\\
	\While {$\mathrm{sum}(\mathbf{x}^\textup{bio}) < F^\textup{lim} $ \textup{ \& } $n^\textup{samp} < n^\textup{site}$} {
		$\mathbf{f}^\textup{com} = \mathrm{sample}(\mathbf{f}^\textup{site}, n^\textup{samp}, \mathbf{p}^\textup{site})$\\
		$\mathbf{x}^\textup{bio} = $ elements of $\mathbf{b}$ with indices $ \in \mathbf{f}^\textup{com}$\\
		$n^\textup{samp} = 1.2n^\textup{samp}$\\
	}
	
	$n^\textup{samp} = $ index of the minimum value of $|\mathrm{cumsum}(\mathbf{x}^\textup{bio}) - F^\textup{lim}|$\\
	$\mathbf{f}^\textup{com} = $ first $n^\textup{samp}$ elements of $\mathbf{f}^\textup{com}$ \\
	$\mathbf{x}^\textup{bio} = $ elements of $\mathbf{b}$ with indices $ \in \{\mathbf{f}^\textup{com}, \mathbf{f}^\textup{surv}\}$ \\
	$\mathbf{x}^\textup{num} = $ rows of $\mathbf{n}$ with row-indices $\in \{\mathbf{f}^\textup{com}, \mathbf{f}^\textup{surv}\}$\\
	
	\Return $\mathbf{x}^\textup{num}$, $\mathbf{x}^\textup{bio}$, $\mathbf{f}^\textup{com}$, $f^\textup{one}$
	\caption{harvest$(\mathbf{b}$, $\mathbf{n}$, $F^\textup{lim}$, $f^\textup{one}$, $\mathbf{f}^\textup{targ}$, $\mathbf{f}^\textup{surv}$, $F^\textup{int}$, $p^\textup{targ} )$}
	
	\label{alg_harvest}
\end{algorithm}

\section{Case study parameters and functions}

The parameters and functions that were used to specify the spatialSim case study scenarios are presented in the following pages that make up this appendix.
\label{app:ompars}

\subsection*{Parameters}

\begin{table}[H]\centering
	\begin{tabular}{ccl}
		\multirow{1}{*}{Process} & \multirow{1}{*}{Parameter} &
		\multicolumn{1}{c}{Value ($\text{OM}_\text{high}$)}  \\ 
		
		\midrule[1 pt]
		\rowcolor{gray!20}
		\multirow{1}{*}{Growth} & $\boldsymbol{\sigma_\delta}$ & $\gsigma$ \\
		
		\rowcolor{gray!10}
		& $\mu_\eta$ & $0$ \\
		\rowcolor{gray!10}
		& $\psi_\eta$ & $\psiEta$ \\
		\rowcolor{gray!10}
		& $\rho_\eta$ & $\rhoEta$ \\
		\rowcolor{gray!10}
		& $\kappa_\eta$ & $\kappaEta$ \\
		\rowcolor{gray!10}
		\multirow{-2}{*}{Natural} & $\tau_\eta$ & $\tauEta$ \\
		\rowcolor{gray!10}
		\multirow{-2}{*}{Mortality} & $\mu_\iota$ & $\muIota$ \\
		\rowcolor{gray!10}		
		& $\rho_\iota$ & $\rhoIota$ \\
		\rowcolor{gray!10}
		& $\sigma_\iota$ & $\sigmaIota$ \\
		\rowcolor{gray!10}
		& $\boldsymbol{\alpha_{M}}$ & $\alphaM$ \\

		\rowcolor{gray!20}		
		& $\boldsymbol{F}^\text{lim}$ & $\Flim$ \\
		\rowcolor{gray!20}
		Fishing & $F^\text{int}$ & $\Fint$ \\
		\rowcolor{gray!20}
		Mortality& $p^\text{targ}$ & $\ptarg$ \\
		\rowcolor{gray!20}
		& $\boldsymbol{f}^\text{one}$ & $\fone$ \\

		\rowcolor{gray!10}
		& $\boldsymbol{r_0}$ & $\rNaught$ \\ 
		\rowcolor{gray!10}
		& $\boldsymbol{h}$ & $\steepness$ \\
		\rowcolor{gray!10}		
		& $\boldsymbol{r}^\text{range}$ & $\rrange$ \\
		\rowcolor{gray!10}		
		& $\boldsymbol{r}^\text{cut}$ & $\rcut$ \\
		\rowcolor{gray!10} 
		& $\mathbf{C}_\epsilon$ & Identity matrix \\
		\rowcolor{gray!10}
		& $\mu_\epsilon$ & $\muEps$ \\
		\rowcolor{gray!10}	
		& $\rho_{\epsilon}$ & $\rhoEps$  \\
		\rowcolor{gray!10}
		& $\kappa_\epsilon$ & $\kappaEps$ \\
		\rowcolor{gray!10}
		\multirow{-9}{*}{Recruitment} & $\tau_\epsilon$ & $\tauEps$ \\

		\rowcolor{gray!20}		
		& $v_B$ & 600  \\ 
		\rowcolor{gray!20}		
		& $v_Y$ & $10$    \\ 
		\rowcolor{gray!20}		
		& $v_P$ & 12   \\
		\rowcolor{gray!20}		
		& $v_L$ & 10   \\
		\rowcolor{gray!20}
		& $v_C$ & 2    \\
		\rowcolor{gray!20}
		& $v_F$ & $557\,433$ \\
		\rowcolor{gray!20}		
		\multirow{-7}{*}{Structure} & $v_S$ & $500$ \\
		\midrule[1 pt]
	\end{tabular}
	\caption{Parameters for $\text{OM}_\text{high}$. Note, these parameter values, with the exception of $F^\text{int}$, are the same for the other operating models. In the case of $\text{OM}_\text{med}$, $F^\text{int} = \Finttwo$, and in the case of $\text{OM}_\text{low}$, $F^\text{int} = \Fintthree$.}
	\label{tab_om}
\end{table}

\subsection*{Functions}

\subsubsection*{Growth increment}
The growth of an individual with size $l \in Q_{c,k}$ was calculated from Faben's von Bertalanffy curve \citep{fab65},
\begin{equation}
\mathcal{F}^\Delta(l; \beta^\Delta_c, l_{c, \infty}, \Delta t) = (1-\exp(-\beta^\Delta_c \Delta t)) (l_{c, \infty} - \bar l_{c,k}),
\end{equation}
where $\boldsymbol{\beta} = \gbeta$ describes the growth rates of S. aequilatera and M. murchisoni individuals, $l_{c, \infty} = \glinf$ are the asymptotic sizes S. aequilatera and M. murchisoni individuals can reach, and $\bar l_{c,k}$ is the midpoint of size-interval $Q_{c,k}$.

\subsubsection*{Size-weight relationship}
The weight of an individual with size $l \in Q_{c,k}$ was calculated from the following exponential growth curve,
\begin{equation}
\mathcal{F}^w(l; \beta^w_{1,c}, \beta^w_{2,c}, \beta^w_{3,c}) = \exp(\beta^w_{1,c} + \beta^w_{2,c} \bar l_{c,k} + \beta^w_{3,c} \bar{l}_{c,k}^{2}),
\end{equation}
where $\boldsymbol{\beta}^w_{1} = (-8.586, -7.599)^\intercal$, $\boldsymbol{\beta}^w_{2} = (0.168, 0.118)^\intercal$, and $\boldsymbol{\beta}^w_{3}=(-0.001, 0.001)^\intercal$.

\subsubsection*{Fishing selectivity}
The probability of a species $c$ individual being retained by fishing gear was 0.95 if they had size $l \in Q_{c,1}$ and 1 otherwise,
\begin{equation}
\mathcal{F}^\zeta(l) = 
	\begin{cases}
		0.95, l \in Q_{c,1} \\
		1.00, l \notin Q_{c,1}.
	\end{cases}
\end{equation}

\subsubsection*{Sexual maturity}
The probability of a species $c$ individual with size $l \in Q_{c,k}$ being sexually mature was calculated from the logistic curve,
\begin{equation}
\mathcal{F}^m(l; l_{50,c}, l_{95,c}) = \frac{1}{1+\exp\left(-\log(19)\frac{\bar{l}_{c,k}-l_{50,c}}{l_{95,c}-l_{50,c}}\right)}.
\end{equation}
Here, the size at which 50\% of individuals were sexually mature was $\boldsymbol{l}_{50} = 0.25\boldsymbol{l}_\infty$ and the size at which 95\% of individuals were sexually mature was $\boldsymbol{l}_{95} = 0.5\boldsymbol{l}_\infty$.

\subsubsection*{Probability of recruitment given environmental conditions}
The probability of a species $c$ individual recruiting conditional on the environmental conditions at each time point and location was specified using two independent components that included (1) an annual seasonality term, $\xi^\text{seas}_t$, and (2) a time-invariant bathymetrical term, $\xi^\text{bath}_{c,s}$,
\begin{equation}
\xi_{c,s,t} = \xi^\text{seas}_t  \xi^\text{bath}_{c,s} 
\end{equation}
The annual seasonality term, $\xi^\text{seas}_t$, was equal to $0$ if $t$ represented one of the first six months of the year and $1/6$ otherwise. The bathymetrical term was calculated from a lognormal distribution,
\begin{equation}
\xi^\text{bath}_{c,s} = \int_{e_s-1}^{e_s} \mathrm{lognorm}(x, \nu^e_{c}, \tau^e_c) \mathrm{d}x,
\end{equation}
where $e_s$ is the depth at location $s$ to the nearest meter and $\mathrm{lognorm}(\cdot)$ is the lognormal density function. The mean and standard deviation parameters on the log scale for \textit{S. aequilatera} and \textit{M. murchisoni} individuals were $\boldsymbol{\nu}^e = (1.32, 1.47)^\intercal$ and $\boldsymbol{\tau}^e= (0.31, 0.27)^\intercal$, respectively.

\end{appendices}


\end{document}